\documentclass[preprint,groupedaddress,floatfix,%
nofootinbib]{revtex4}

\usepackage{euscript}
\usepackage{dcolumn}
\usepackage{graphicx}
\usepackage{epsfig}
\usepackage{hyperref}
\usepackage{graphicx}
\usepackage{dcolumn}
\usepackage{longtable}
\usepackage{times}
\usepackage{amsmath,amssymb,bm}
\usepackage[english]{babel}
\usepackage[latin1]{inputenc}
\usepackage{times}
\usepackage[T1]{fontenc}
\usepackage{color}
\usepackage{fancybox}
\usepackage{sidecap}
\usepackage{graphicx}
\usepackage{epsfig}
\usepackage{psfrag}
\usepackage{bbm}
\usepackage[english]{babel}
\usepackage[latin1]{inputenc}
\usepackage{times}
\usepackage[T1]{fontenc}
\usepackage{color}
\usepackage{fancybox}
\usepackage{sidecap}
\usepackage{psfrag}
\usepackage{bbm}
\usepackage{wasysym}
\usepackage{slashed}
\usepackage{tikz}
\usepackage{euscript}
\usepackage{CJK}

\newcommand{\mathi}{\mathrm{i}}
\newcommand{\mathe}{\mathrm{e}}

\begin{document}

\title{A Refined Algorithm For the EPR model}
\author{Wenxuan Tao}
\author{Fen Zuo\footnote{Email: \textsf{zuofen@miqroera.com}}}
\affiliation{Shanghai MiQro Era Digital Technology Co. Ltd., Shanghai, China}

\begin{abstract}
The Einstein-Podolsky-Rosen~(EPR) model is an analogous model of the anti-ferromagnetic Heisenberg model or the equivalent quantum maximum-cut problem, proposed by R. King two years ago. Adjacent qubits in the model prefer symmetric EPR/Bell parings rather than the antisymmetric one, in order to maximize the energy. Recently, two groups independently develop specific algorithms for the highest-energy state with approximation ratio $\frac{1+\sqrt{5}}{4}\approx.809$, based on maximum fractional matchings. Here we try to refine one of the two algorithms by devising homogeneous/quasi-homogeneous fractional matchings, with the aim to distribute quantum entanglement as much as possible. For regular graphs $G_d$, we immediately obtain increasing approximation ratios $r_d$ with $r_2=\frac{3+\sqrt{5}}{6}\approx.872$. For irregular graphs, we show such a refinement could still guarantee nice performance if the fractional matchings are chosen properly.

\end{abstract}
 \maketitle

\tableofcontents

\section{The EPR model}

We start with the anti-ferromagnetic Heisenberg model, which is equivalent to the so-called Quantum Maximum Cut~(QMC) problem. In the latter problem, the Hamiltonian is defined on a weighted graph $G(V,E,w)$:
\begin{equation}
H^\textrm{QMC}_G= \sum_{(ij)} w_{ij}h_{ij}\equiv\frac{1}{2}\sum_{(ij)} w_{ij}(I_iI_j-X_iX_j-Z_iZ_j-Y_iY_j),
\end{equation}
where the summation is over all edges $(ij)\in E_G$, and all the weights are supposed to be positive. The aim is to maximize the corresponding energy, or the expectation value of the Hamiltonian.  Thus adjacent qubits would favor the antisymmetric pairing
\begin{equation}
|\psi_-\rangle=\frac{1}{2}(|01\rangle-|10\rangle).
\end{equation}
The Einstein-Podolsky-Rosen (EPR) model could be considered as the symmetric counterpart~\cite{King-2023} of QMC.
 Without loss of generality, the Hamiltonian could be taken as
 \begin{equation}
H^\textrm{EPR}_G= \sum_{(ij)} w_{ij}g_{ij}\equiv\frac{1}{2}\sum_{(ij)} w_{ij}(I_iI_j+X_iX_j+Z_iZ_j-Y_iY_j).
\end{equation}
Then adjacent qubits would favor the symmetric pairing
\begin{equation}
|\phi_+\rangle=\frac{1}{2}(|00\rangle+|11\rangle),
\end{equation}
in order to maximize the energy.
According to the analyses in~\cite{PM-2015}, QMC is QMA-hard and EPR is in StoqMA~\cite{Bravyi-2006}.

With the so-called monogamy of entanglement argument~\cite{CKW-1999}, the maximum eigenvalue of both QMC and EPR could be bounded by maximum-weight fractional matching~(MWFM) as~\cite{AGM-2020}\cite{Lee-2024}:
\begin{equation}
\lambda_\textrm{max}(H_G)\le w_G+w(\textrm{FM}_G). \label{eq.FMb}
\end{equation}
Here $w_G$ is total weight, and $w(\textrm{FM}_G)$ is the value of the MWFM $\textrm{FM}_G$. Possible strengthening of the bound to maximum-weight matching is discussed in \cite{GSS-2025}\cite{APS-2025}.

In the rest of the paper, we only consider the EPR model, and more specifically the highest-energy state of the EPR model.

\section{THE ALMPS Algorithm}

The above upper bound suggests that one could devise nice approximation algorithms based on MWFM. This is the strategy adopted recently in both \cite{JN-2025} and \cite{ALMPS-2025}. They independently develop approximation algorithms for EPR with the same approximation ratio $r_0\approx.809$, but the implementation are quite different.
The algorithm in \cite{JN-2025} uses a heavy mixed state, which is not quite convenient to generalize. So we focus on the approach of \cite{ALMPS-2025}, and call it ALMPS for short. The ALMPS algorithm utilizes a nice state proposed in \cite{King-2023}, which we now review.

\subsection{Magic Graph States}

In the measurement-based quantum computing scenario, the so-called graph states are indispensable, which are defined as
\begin{equation}
|\psi_G\rangle \equiv\prod_{(ij)}CZ_{ij} |+\rangle^{\otimes n}.
\end{equation}
Since only $CZ$ and $H$ gates are involved, graph states are necessary Clifford. Ref.~\cite{King-2023} introduces generalized graph states as
\begin{equation}
|\chi\rangle \equiv\prod_{(ij)}\mathe^{\mathi \theta_{ij}P_iP_j} |0\rangle^{\otimes n},
\end{equation}
with
\begin{equation}
P=T Y T^\dagger =\frac{1}{\sqrt{2}}(X-Y),\quad Q=T X T^\dagger=\frac{1}{\sqrt{2}}(X+Y).
\end{equation}
Here all the two-qubit gates $\mathe^{\mathi \theta_{ij}P_iP_j}$ also communicate with each other as the $CZ$ gates above, making the states well defined. Moreover, now $|\chi\rangle$ necessarily involve a huge number of $T$ gates, so we may call them  magic graph states. Due to the inclusion of magic, we could expect them to possess strong expressive power.

With the above definitions of $P$ and $Q$ gates, we may re-express the terms involving $X$ and $Y$ in the EPR Hamiltonian as
\begin{equation}
X_iX_j-Y_iY_j=P_iQ_j+Q_iP_j.
\end{equation}
So to compute the energy, we need to calculate the expectation values of $Z_iZ_j$, $P_iQ_j$ and $Q_iP_j$ on $|\chi\rangle$. All these have been done in detail in \cite{King-2023}, with the concrete expressions:
\begin{eqnarray}
\langle\chi| Q_iP_j|\chi\rangle &=&\sin 2\theta_{ij} \prod_{k\in K} \cos 2\theta_{ik},\label{eq.QP}\\
\langle\chi| P_iQ_j|\chi\rangle &=&\sin 2\theta_{ij} \prod_{l\in L} \cos 2\theta_{jl},\label{eq.PQ}\\
\langle \chi| Z_iZ_j|\chi\rangle &=&\sum_{S\subset T, |S|\textrm{even}} \prod_{s\in S}\sin 2\theta_{is}\sin 2\theta_{sj} \prod_{k\in K\setminus S} \cos 2\theta_{ik}\prod_{l\in L\setminus S} \cos 2\theta_{lj}.\label{eq.ZZ}
\end{eqnarray}
Here
\begin{equation}
K\equiv \{k\in V_G: k\sim i, k\ne j\}, \quad L\equiv \{l\in V_G: l\sim j, l\ne i\},
\end{equation}
and
\begin{equation}
T=\{t\in V_G: t\sim i, t\sim j\}.
\end{equation}
One can see that only the expression of $\langle \chi| Z_iZ_j|\chi\rangle$ inherits some quantumness, and may not be efficiently calculated with classical methods.

With these explicit expressions, one may be tempted to directly variate all the parameters $\{\theta_{ij}\}$ to get higher energies. Since the number of parameters is so huge, this would not be an easy task anyhow. Even this could be done efficiently, it would not necessarily guarantee a very nice approximation. In fact, we have the following claim:

\textbf{Claim 1.} Variation of magic graph states $|\chi\rangle$ gives an approximation ratio for EPR no larger than $r_2=\frac{3+\sqrt{5}}{6}\approx .872$.

This can be easily confirmed by considering $G=K_{2,2}$. It shows that, the power of the magic graph states $|\chi\rangle$ in solving EPR has an obvious limitation. So instead of devising universal algorithms for all graphs, one could consider an alternative strategy and develop specific algorithms for specific graphs. This will be elucidated in detail later.

Still, such a brute-force approach gives very nice results for many graphs, with approximation ratios often much larger than $r_2$. So we boldly make the following conjecture:

\textbf{Conjecture 1.} Variation of magic graph states $|\chi\rangle$ could achieve the approximation ratio $r_2$ for EPR.

We will give some evidences for this later.

\subsection{The ALMPS Algorithm}

The ALMPS algorithm considers more sophisticated choices for the parameters~\cite{ALMPS-2025}. From the expressions (\ref{eq.QP},\ref{eq.PQ},\ref{eq.ZZ}) one can see that the energy on an individual edge is affected by the parameters on the adjacent edges. \cite{ALMPS-2025} uses MWFM to cleverly convert these parameters back to the one on the current edge. More precisely, they choose the parameters $\theta_{ij}$ according to the matching fraction $m_{ij}$ in MWFM as
\begin{equation}
\cos 2\theta_{ij}=\exp[-\kappa\cdot m_{ij}],
\end{equation}
where $\kappa$ is a positive constant, and $\theta_{ij}$ is restricted to be in $[0,\pi/4]$. Now if we abandon the higher powers in (\ref{eq.ZZ}), we get the following estimation of the individual energy term:
\begin{equation}
\langle \chi| g_{ij}|\chi\rangle \ge T(\kappa,m_{ij})\equiv \frac{1}{2}(1+\exp[-2\kappa(1-m_{ij})]+2\sqrt{1-\exp[-2\kappa m_{ij}]}\exp[-\kappa(1-m_{ij})]).
\end{equation}
Since the right-hand side depends only on $m_{ij}$, we could easily bound it with $m_{ij}$ as we want. Thus we define
\begin{equation}
R(\kappa,m_{ij})\equiv T(\kappa,m_{ij})/(1+m_{ij}),
\end{equation}
and try to maximize it for all the values that $m_{ij}$ may take.

According to \cite{Matching-1986}, every vertex of the fractional matching polytope take the basic form, which consists of disconnected edges and odd cycles. Therefore, MWFM of the basic form generally contains matching fractions $0,1/2,1$. In the degenerate case, other values could also appear. Therefore, one could consider the following maximum-minimum problem:
\begin{equation}
r_{0}\equiv\max_{\kappa>0} \min _{x\in[0,1]} R(\kappa,x).\label{eq.r0}
\end{equation}
\cite{ALMPS-2025} proves that it is half the golden ratio $\varphi= \frac{1+\sqrt{5}}{2}$:
\begin{equation}
r_{0}=\frac{\varphi}{2}\approx .809,
\end{equation}
and the corresponding parameter is
\begin{equation}
\kappa_{0}\equiv \frac{1}{2}\ln \varphi \approx .240.
\end{equation}
With the parameter $\kappa$ fixed this way, one could employ the upper bound (\ref{eq.FMb}) to show
\begin{eqnarray}
\frac{\langle \chi| H_G^\textrm{EPR} |\chi\rangle}{\lambda_\textrm{max}(H_G^\textrm{EPR})}&\ge& \frac{\langle \chi| H_G^\textrm{EPR} |\chi\rangle}{w_G+w(\textrm{FM}_G)}\nonumber\\
&=&\frac{\sum_{(ij)} w_{ij}\langle \chi| g_{ij}|\chi\rangle}{\sum_{(ij)} w_{ij}(1+m_{ij})}\nonumber\\
&\ge & \min_{x\in[0,1]} R(\kappa_0,x)=r_0.\label{eq.ALMPS}
\end{eqnarray}
Therefore, the ALMPS algorithm has an approximation ratio
\begin{equation}
r(\textrm{ALMPS})\ge r_{0}\approx .809.
\end{equation}
A simple analysis shows that, the appearance of both the matching fractions $0$ and $1$ forces the ratio to be no larger than $r_0$. This is emphasized recently in \cite{APS-2025}. Since the basic form MWFM generally contains unmatched and full-matched edges, if we want to go beyond $r_0$ we should turn to degenerate fractional matchings, or even non-maximum ones.

\section{A Refined Algorithm}

From the above analysis of the ALMPS algorithm, we realize that in order to increase the ratio we must shrink the range of the matching fractions. The matching fractions in the ALMPS algorithm further determine the entanglement degrees on the edges. So when we constrain the matching fractions, we will re-distribute quantum entanglement, making it more homogeneous across the graph. Therefore, we call the refined algorithm ``Fractional Entanglement Distribution'', FED for short. The implementation of FED is almost identical to ALMPS, and the only difference is the choice of the fractional matching. Due to this, FED is just a refinement of ALMPS.

Of course we still need to keep the matching value large enough when we constrain the matching fractions. If we insist on MWFMs, we should turn to degenerate solutions. That is, the optimal solutions of the corresponding linear program appear not just on vertices, but also on lines and surfaces. This is quite common in un-weighted graphs. We will first consider the situation when the matching faction takes a single value, with regular graphs the typical cases.

\subsection{Regular graphs}

For $d$-regular graphs $G_d$, there is a natural perfect fractional matching, in which every edge has the matching fraction $1/d$. We call this a homogeneous Fractional Matching, hFM for short. With this choice the maximum-minimum problem in ALMPS reduces to
\begin{equation}
r_d=\max_{\kappa>0} R(\kappa,1/d).
\end{equation}
It is very easy to obtain the solutions for small degrees, as shown in Tab. \ref{Tab.Gd}.

\begin{table}[h]
\begin{center}
\begin{tabular}{|c|c|c|}
\hline
$\quad d \quad $  &  $\quad r_d \quad $ &   $\quad \kappa_d \quad $          \\
\hline\hline
2           &  .872  &  .324    \\
\hline
3        &  .894 &  .203    \\
\hline
4       &  .912 &  .147   \\
\hline
5         &  .924 &  .115   \\
\hline
6       &  .934 &  .0945    \\
\hline
7        &  .942 &  .080  \\
\hline
8         &  .948 &  .0692   \\
\hline
9     &  .953 &  .061 \\
\hline
10      &  .957 &  .0544\\
\hline
\end{tabular}
\\
\caption{Approximation ratios ~$r_d$ of FED on regular graphs~$G_d$, together with the parameters~$\kappa_d$.}
\label{Tab.Gd}
\end{center}
\end{table}

 Notice $r_2$ has been defined and used in previous sections. It seems that as the degree $d$ increases, the ratios $r_d$  increases monotonously. We believe this behavior holds up to infinity. We numerically confirm this up to $d=50$.  Repeating the analysis of ALMPS (\ref{eq.ALMPS}), we immediately get

\textbf{Theorem 1.} For un-weighted $d$-regular graphs $G_d$ with $d\ge 2$, FED achieves an approximation ratio at least $r_d$. Moreover, $r_d$ increases monotonously with $d$.

Since $r_d$ increases with $d$, and EPR defined on $G_1$ is trivial, we also have

\textbf{Corollary 1.} Conjecture 1 holds for all un-weighted regular graphs.

\subsection{Irregular graphs}

Of course un-weighted regular graphs are very special. When weights are added, the degeneracy of MWFMs often disappears. And in irregular graphs, the above homogeneous matching will not exist anymore. However, we hope that the above strategy would still benefit, if we cleverly devise our matching. Since we only consider positive weights, we could use some tricks to properly convert them into integers. Then we could split the edge with integral weights into multiple un-weighted edges. So we focus on un-weighted graphs, possibly multi-graphs, but with no self-loops.

We start with a general fractional matching. Let the minimum fraction to be $1/\Delta$, and the maximum one to be $1/\delta$. So the matching fractions take values from the interval $I_{\delta,\Delta}=[\Delta^{-1},\delta^{-1}]$. When $I_{\delta,\Delta}$ is strictly smaller that $[0,1]$, we can not guarantee that the matching would be maximum. However, we could still manage to get the optimal solution $\textrm{FM}_G^{\delta,\Delta}$ with the constraint $m_e\in I_{\delta,\Delta}$. Then we could calculate the ratio
\begin{equation}
s^{\delta,\Delta}_G\equiv \frac{|\textrm{FM}^{\delta,\Delta}_G|}{|\textrm{FM}_G|},
\end{equation}
to see whether it is close enough to the maximum matching. In fact, what really matters in the algorithms is the sum of the total weight and the matching value, so we need to analyse the following shifted ratio
\begin{equation}
\hat s^{\delta,\Delta}_G\equiv \frac{|E_G|+|\textrm{FM}^{\delta,\Delta}_G|}{|E_G|+|\textrm{FM}_G|}.
\end{equation}
Obviously $\hat s^{\delta,\Delta}_G$ would be closer to $1$ than the original $s^{\delta,\Delta}_G$. We hope we could make a complete characterization of these ratios in the future.

Now we input this matching $\textrm{FM}_G^{\delta,\Delta}$ into our algorithm. We would be led to the following maximum-minimum problem
\begin{equation}
r_{\delta,\Delta}=\max_{\kappa>0} \min _{x\in[\Delta^{-1},\delta^{-1}]} R(\kappa,x). \label{eq.rdD}
\end{equation}
This is a direct generalization of the original expression (\ref{eq.r0}). Now repeating the analysis of ALMPS (\ref{eq.ALMPS}), we get

\textbf{Theorem 2.} For an un-weighted graph $G$, if we choose the fractional matching $\textrm{FM}_G^{\delta,\Delta}$ with fraction interval $I_{\delta,\Delta}$ and shifted ratio $\hat s^{\delta,\Delta}_G$, then the FED algorithm guarantees an approximation ratio $r_{\delta,\Delta}\cdot \hat s^{\delta,\Delta}_G$.

It is not difficult to see that, if we take $I_{\delta,\Delta}=[0,1]$, then we can achieve $\hat s^{\delta,\Delta}_G=1$, and meanwhile $r_{\delta,\Delta}=r_0$, so we recovers the original ALMPS algorithm. When we take $\delta=\Delta=d$, then $r_{d,d}=r_d$; and if $G$ is regular, we have $\hat s^{d,d}_{G_d}=1$, thus we reproduce all the results in the previous subsection. So the above theorem actually embraces the ALMPS algorithm and the specific results for regular graphs. Besides these two extreme cases, could the theorem still be beneficial? Below we give a concrete construction to show that, it is indeed so.

Given a graph $G$, possibly a multi-graph, we define the edge degree as
\begin{equation}
d_{ij}=\max\{d_i,d_j\}.
\end{equation}
To see the distribution of the edge degree, we further define the minimum and maximum edge degree as
\begin{equation}
\delta=\min\{d_e|e\in E_G\},\quad \Delta=\max\{d_e|e\in E_G\}.
\end{equation}
Notice that we have abused the notations slightly here, but we will soon see that they are consistent. So the edge degree takes values in the interval $[\delta,\Delta]$. Since the edge degree is induced by the vertex degree, this interval would be contained in the interval of vertex degrees. More specifically, $\Delta$ would coincide with the maximum vertex degree, while $\delta$ will in general exceed the minimum vertex degree. As a result, the distribution of edge degrees would be more concentrated. A typical example is the complete bipartite graph $K_{l,m}$, whose vertex degrees take values $\{l,m\}$, while the edge degree takes a single value $\max\{l,m\}$. When the minimum edge degree coincides with the maximum one, that is when $\delta=\Delta=d$, we call the graph edge-degree $d$-regular.

\begin{figure}[h]
\centering
	\includegraphics[width=0.35\textwidth]{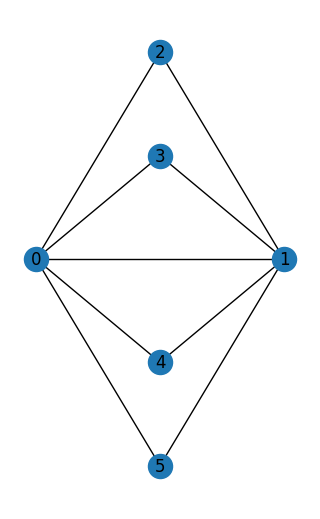}
\caption{Counter example: Quasi-homogeneous fractional matchings of edge-degree regular graphs need not be maximum.}\label{fig.triangle}
\end{figure}

Now we use edge degree to define a simple fractional matching
\begin{equation}
m_{ij}=1/d_{ij}, \quad \forall (ij)\in E_G.
\end{equation}
We call it a quasi-homogeneous Fractional Matching, qhFM for short. Obviously this is a direct generalization of the homogeneous Fractional Matching for regular graphs. It is not difficult to check that it is indeed a feasible fractional matching. Here we want to point out that, even for edge-degree regular graphs, qhFM needs not be maximum. A counter example is given in Fig.\ref{fig.triangle}. The quasi-homogeneous matching is $9/5$, while the maximum matching is $2$.

Now since edge degrees take values in $[\delta,\Delta]$, matching fractions take values in $I_{\delta,\Delta}=[\Delta^{-1},\delta^{-1}]$. We could apply Theorem 2 directly. Let us calculate some examples to see how it works. All the graphs considered are very small, so we could replace the matching interval $I_{\delta,\Delta}$ in Eq.(\ref{eq.rdD}) with the explicit set of matching fractions.

\textbf{Example 1.} Consider the complete bipartite graph $K_{3,6}$. We have $\delta=\Delta=6$, so $r_{6,6}=r_6\approx.934$. qhFM is actually maximum here, so $\hat s^{6,6}_G=1$. Therefore, FED has an approximation ratio at least $.934$, much better than $r_0$ and $r_2$.

\textbf{Example 2.} For the graph in fig.~\ref{fig.triangle}, $\delta=\Delta=5$, so $r_{5,5}=r_5\approx.924$. Now $|\textrm{qhFM}_G|=9/5$, $|\textrm{FM}_G|=2$, so $\hat s^{5,5}_G=.982$. Thus FED has approximation ratio no less than $r_5\cdot \hat s^{5,5}_G\approx .907$, still better than $r_0$ and $r_2$.

\begin{figure}[h]
\centering
	\includegraphics[width=0.6\textwidth]{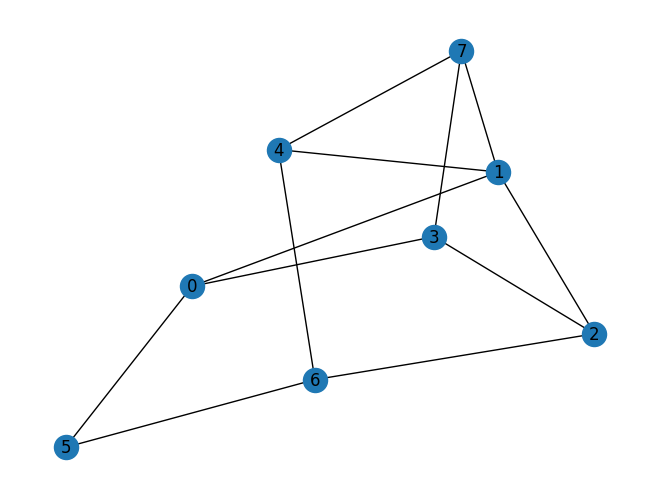}
\caption{An example with edge degree range $[3,4]$. }\label{fig.example3}
\end{figure}

\textbf{Example 3.} For the graph in fig.~\ref{fig.example3}, $\delta=3$, $\Delta=4$, $r_{3,4}\approx r_3\approx .894$. And $\hat s^{3,4}_G\approx .979$. So FED has approximation ratio at least $r_3\cdot\hat s^{3,4}_G\approx .8755$, also better than $r_0$ and $r_2$.

\begin{figure}[h]
\centering
	\includegraphics[width=0.45\textwidth]{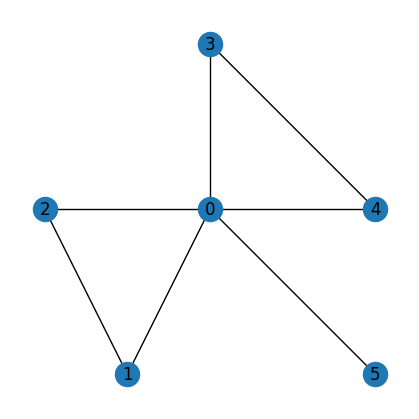}
\caption{An example where the quasi-homogeneous fractional matching fails to induce a nice approximation for EPR.}\label{fig.example4}
\end{figure}

Of course, qhFM is a rather rude construction. One may easily devise specific graphs so that qhFM fails to induce a nice approximation ratio for EPR. Here is an example:

\textbf{Example 4.} For the graph in fig.~\ref{fig.example4}, $\delta=2$, $\Delta=5$, $r_{2,5}\approx r_2\approx.872$. And $|\textrm{qhFM}_G|=2$, $|\textrm{FM}_G|=3$ , so $\hat s^{2,5}_G= .9$. We get an approximation ratio $.785$, lower than $r_0$.

This can be easily improved by choosing a general fractional matching in a more sophisticated way.

\textbf{Example 4'.} Again for the graph in fig.~\ref{fig.example4}, take $\Delta=5$, but $\delta=1.25$, we get $|\textrm{FM}_G^{1.25,5}|=2.6$, so $\hat s^{1.25,5}_G= .96$. Also $r_{1.25,5}\approx.868$. This results in an approximation ratio $.833$, better than $r_0$. It is still lower than $r_2$, and one may wonder if it serves as a counter example to Conjecture 1. This is not true, since here the fractional matching bound overestimates the highest energy by about $10\%$.

We leave the general fractional matchings for future study.

\section{Discussion}
In this paper we have refined the ALMPS algorithm by choosing more general fractional matchings. Two criterions are considered when making such choices: the matching fractions should be concentrated enough, and the matching value should be large enough. When the matching fractions are concentrated, the distribution of quantum entanglement would be more homogeneous, and we will obtain larger quantum contributions in the total energy. This is most significant for regular graphs, where we get rather nice approximation ratios $r_d$. Even for general graphs, if we choose the fractional matching properly, this approach still guarantees nice performance no worse than ALMPS.

However, both the ALMPS algorithm and the refinement here work within the space of the specific magic graph states. As we point out previously, magic graph states deteriorate on $K_{2,2}$ and general $2$-regular graphs. More specifically, magic graph states fail to provide nice approximations to EPR on individual cycles.  We speculate that this is due to the fact that only communicative gates are used in constructing these states. Using only communicative gates enables us to completely neglect the time order, but also makes the states very special. To overcome the limitations of magic graph states, it seems that general quantum circuits with definite operational orders are inevitable.

It is also worthy to mention that, Heisenberg model defined on cycles can be exactly solved by using Bethe ansatz~\cite{Bethe-1931}.
Perhaps integrability could give some hints on how to approximate EPR on cycles with quantum circuits. We would like to investigate this in the future.

\section*{Acknowledgments}
We thank Xiaxia Guan for helpful discussions on fractional matchings and for sharing her paper ``Fractional matchings on regular graphs''.

%


\end{document}